\documentclass[conference]{IEEEtran}
\usepackage{amsmath}
\usepackage{graphicx}
\usepackage{hyperref}
\usepackage{tikz}
\usetikzlibrary{shapes.geometric, arrows, positioning, fit, backgrounds}
\usepackage{pdfpages}

\ifCLASSINFOpdf

\else

\fi

\begin{document}

\title{Poster: Privacy-Preserving Compliance Checks on Ethereum via Selective Disclosure}

\author{\IEEEauthorblockN{Supriya Khadka}
	\IEEEauthorblockA{Coventry University\\
		khadkas25@uni.coventry.ac.uk}
    \and
	\IEEEauthorblockN{Dhiman Goswami}
	\IEEEauthorblockA{George Mason University\\
		dgoswam@gmu.edu}
	\and
	\IEEEauthorblockN{Sanchari Das}
	\IEEEauthorblockA{George Mason University\\
		sdas35@gmu.edu}
        }


\maketitle

\begin{abstract}
Digital identity verification often forces a privacy trade-off, where users must disclose sensitive personal data to prove simple eligibility criteria. As blockchain applications integrate with regulated environments, this over-disclosure creates significant risks of data breaches and surveillance. This work proposes a general Selective Disclosure Framework built on Ethereum, designed to decouple attribute verification from identity revelation. By utilizing client-side zk-SNARKs, the framework enables users to prove specific eligibility predicates without revealing underlying identity documents. We present a case study, \emph{ZK-Compliance}, which implements a functional \emph{Grant, Verify, Revoke} lifecycle for age verification. Preliminary results indicate that strict compliance requirements can be satisfied with negligible client-side latency ($< 200$ ms) while preserving the pseudonymous nature of public blockchains.
\end{abstract}


%
\IEEEpeerreviewmaketitle

\section{Motivation}
Public blockchains operate on the principle of radical transparency to ensure trustless verification of value transfer \cite{conti2018survey}. While this architecture ensures data integrity and auditability, it creates a hostile environment for Personally Identifiable Information (PII) \cite{buterin2016privacy}. As the blockchain ecosystem expands into regulated sectors, the conflict between this transparency and the need for user privacy has become a critical bottleneck.

The fundamental issue is that data placed on a public blockchain is immutable and globally visible. Although networks like Bitcoin provide a degree of pseudonymity, research has demonstrated that simple clustering heuristics can easily link on-chain behavior to real-world identities~\cite{ron2013quantitative, maesa2016uncovering,agarwal2025systematic}. This issue is intensified as decentralized applications (dApps) integrate with regulated financial markets. To comply with Know Your Customer (KYC) and Anti-Money Laundering (AML) regulations, protocols often require users to verify their identities~\cite{fatf2019vasp}. Under current paradigms, this creates a \textit{transparency trap}: participation in a compliant economy results in the persistent linkage of a verified real-world identity with an immutable transaction history~\cite{buterin2016privacy}. 


To address these concerns, the industry currently relies on centralized verification models where users upload documents to a Trusted Third Party (TTP)~\cite{belchior2021survey}. However, this reintroduces the centralization risks that blockchain technology aims to eliminate~\cite{bruhner2023bridging}. Centralized points of failure are prime targets for attacks, and history includes numerous instances of custodians suffering catastrophic breaches~\cite{conti2018survey}. Furthermore, reliance on intermediaries grants third parties the power to censor users, failing to provide a truly sovereign solution~\cite{buterin2016privacy}.

In response to these limitations, this work contributes a \textbf{Selective Disclosure Framework} designed to address the structural tension between regulatory compliance and user privacy on public blockchains. By utilizing Zero-Knowledge Proofs (ZKPs), specifically zk-SNARKs, our framework enables attributes to be verified on-chain without exposing the underlying identity data. This approach effectively decouples identity storage from verification logic, mitigating the risks of both the \textit{transparency trap} and centralized data breaches.

\section{System Architecture}
The proposed framework is designed around a \textit{client-side proving} model, ensuring that raw identity data never leaves the user's device. The system coordinates interactions between the User (Prover), a trusted Issuer (e.g., a government agency), and a Service dApp (Verifier) through a set of on-chain and off-chain components.

\subsection{Core Components}
The architecture consists of four primary components:
\begin{itemize}
    \item \textit{Identity Vault (Client/Browser)}: A local secure storage mechanism that holds the user's raw attributes (e.g., birthdate) in an encrypted format, ensuring PII remains strictly non-custodial.
    \item \textit{ZK Prover (SnarkJS)}: A client-side module running inside the browser that generates proofs using the user's private inputs and a cryptographic salt.
    \item \textit{Verifier Contract}: An auto-generated Solidity contract that validates the cryptographic proof on-chain without viewing the inputs.
    \item \textit{Access Registry}: A \textit{state manager} contract that records valid access grants and manages the revocation lifecycle.
\end{itemize}

\begin{figure}[ht]
    \centering
    \resizebox{\linewidth}{!}{%
    \begin{tikzpicture}[
        node distance=1.5cm and 1cm,
        auto,
        every node/.style={font=\sffamily\small},
        process/.style={rectangle, minimum width=2.2cm, minimum height=1cm, align=center, draw=black, fill=white, rounded corners, thick},
        secret/.style={rectangle, minimum width=2.2cm, minimum height=1cm, align=center, draw=red!80!black, dashed, fill=red!5, thick},
        contract/.style={rectangle, minimum width=2.2cm, minimum height=1cm, align=center, draw=blue!80!black, fill=blue!5, thick},
        arrow/.style={thick,->,>=stealth, rounded corners},
        labelbox/.style={font=\bfseries\footnotesize, text=gray!70!black, align=left}
    ]

    \node (user) [secret] {User Secrets\\(Birthdate + Salt)};
    \node (prover) [process, right=1.1cm of user] {ZK Prover\\(Browser)};
    \node (proof) [process, right=1.6cm of prover] {ZK Proof\\(Encrypted)};

    \node (verifier) [contract, below=1.5cm of proof] {Verifier\\Contract};
    \node (registry) [contract, left=1.5cm of verifier] {Access\\Registry};

    \node (dapp) [process, below=1.5cm of registry, fill=purple!5, draw=purple!80!black] {Service dApp\\(Consumer)};


    \draw [arrow, blue!80!black] (user) -- node[above, scale=0.8] {1. Input} (prover);
    \draw [arrow, blue!80!black] (prover) -- node[above, scale=0.8] {2. Generate} (proof);
    \draw [arrow, blue!80!black] (proof) -- node[right, scale=0.8] {3. Transact} (verifier);
    \draw [arrow, blue!80!black] (verifier) -- node[above, scale=0.8] {4. Register} (registry);

    \draw [arrow, purple!80!black, dashed] (dapp.east) to[bend right=45] node[right, scale=0.8, pos=0.5] {5. Query} (registry.south east);
    \draw [arrow, purple!80!black, dashed] (registry.south west) to[bend right=45] node[left, scale=0.8, pos=0.5] {6. Grant} (dapp.west);

    \draw [arrow, red!80!black, thick] (user) -- node[left, scale=0.8, pos=0.6, xshift=-12pt, align=right] {7. Revoke\\(Kill Switch)} (registry);

    \begin{pgfonlayer}{background}
        \node [fit=(user) (prover) (proof), draw=gray!30, fill=gray!5, rounded corners, dashed, inner sep=0.3cm] (clientbox) {};
        \node [above right, xshift=-5pt] at (clientbox.north west) [labelbox] {Client Side (Off-Chain)};

        \node [fit=(verifier) (registry), draw=blue!20, fill=blue!5, rounded corners, inner sep=0.3cm] (chainbox) {};
        \node [above right, xshift=5pt] at (chainbox.north west) [labelbox, text=blue!70!black] {Ethereum (On-Chain)};
    \end{pgfonlayer}

    \end{tikzpicture}
    }
    \caption{\textbf{``Grant, Verify, Revoke''} lifecycle. (1-4) User grants access via a ZK proof. (5-6) dApp consumes the access. (7) User can revoke permission at any time.}
    \label{fig:architecture_lifecycle}
\end{figure}
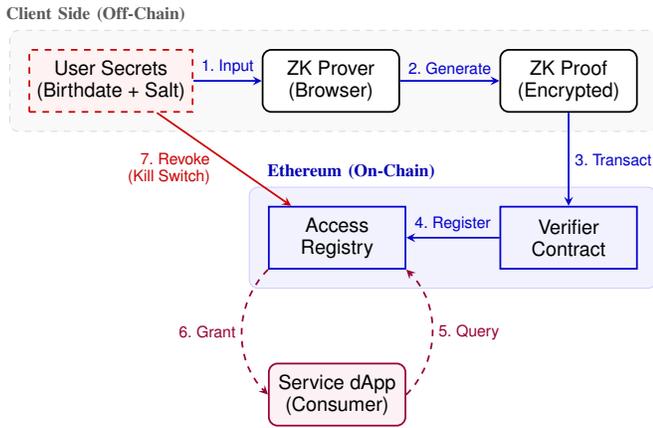

\subsection{The Execution Lifecycle}
The framework coordinates interactions through a structured \emph{Grant, Verify, Revoke} lifecycle, as shown in Figure~\ref{fig:architecture_lifecycle},  that transforms private identity attributes into a public, revocable access grant:

\textit{Phase 1: Grant (Proof Generation)}: The lifecycle begins when a Verifier requests eligibility (e.g., $Age > 18$) for a specific duration. The user's client retrieves the private attribute from the Identity Vault, mixes it with a cryptographic salt to prevent rainbow table attacks, and inputs it into the ZK Prover. The client generates a zk-SNARK proof locally and submits a \texttt{grantAccess} transaction to the Access Registry. The Registry automatically validates the cryptographic proof on-chain; if valid, it mints an \texttt{AccessRecord} representing the user's consent.

\textit{Phase 2: Verify (Access Consumption)}: Once granted, verification becomes a lightweight, constant-time lookup. The Service dApp queries the Registry’s \texttt{validateAccess} function. Instead of processing heavy cryptography, the dApp simply verifies that a valid, non-expired \texttt{AccessRecord} exists for the user. This decouples the computational cost of zero-knowledge proofs from the frequency of user sessions, allowing for repeated verification without repeated transactions.

\textit{Phase 3: Revoke (Sovereign Control)} Unlike static credentials, this architecture provides a stateful kill switch. The user can trigger the \texttt{revokeAccess} transaction at any time to delete their \texttt{AccessRecord}. This instantly invalidates the dApp's ability to verify eligibility, restoring the user's privacy boundary immediately and preventing indefinite data retention.

\section{Preliminary Results: ZK-Compliance}
To validate the feasibility of this architecture, we developed \emph{ZK-Compliance}, a proof-of-concept implementation focused on Age Verification.

\subsection{Implementation Stack}
Written in Circom, the circuit performs a range check ($Year - BirthYear \ge Threshold$) and squares the random salt to enforce entropy. We utilized SnarkJS to enable proof generation within a standard web browser. Smart contracts were deployed on the Ethereum Sepolia testnet, utilizing the BN128 elliptic curve for efficient on-chain pairing checks.

\subsection{Performance \& Cost Analysis}
We evaluated the system across two critical dimensions: user latency and economic viability.
On standard consumer hardware, the browser-based prover generates valid proofs in $< 200$ ms. This indicates that client-side zk-SNARKs introduce negligible friction to the user experience. 

Verification on Ethereum Mainnet, however, is cost-prohibitive: the on-chain Groth16 pairing check consumes $\approx 240,512$ gas, costing $\approx \$15.00$ per session (assuming 20 gwei at $\$3,000$ ETH). In contrast, deploying the Registry on Layer 2 scaling solutions (e.g., Arbitrum) reduces the cost to $< \$0.50$, making the system economically viable for high-volume applications.

\section{CONCLUSION AND FUTURE WORK}
This work demonstrates that regulatory compliance on public blockchains does not require the \textit{transparency trap} of over-disclosure. By using client-side zk-SNARKs, our Selective Disclosure Framework allows users to satisfy strict eligibility predicates while maintaining pseudonymity and sovereign control. The current browser-based prototype provides early evidence that this architecture is capable enough for consumer applications, while restoring the user's ability to \emph{Grant, Verify, and Revoke} access without permanent data retention.

In the future, to mitigate the risk of mempool front-running, we will update the circuit design to cryptographically bind proofs to the user’s \texttt{msg.sender} address. We will also address the risk of deanonymization via cross-application clustering by implementing dApp-scoped nullifiers, ensuring that reusable identities do not create a global correlation graph. Finally, acknowledging the security risks of standard web environments, we aim to investigate isolating proving logic within secure enclaves and exploring off-chain revocation signaling to ensure the on-chain \emph{Grant} history does not leave a persistent inferential trail.

\section*{Acknowledgment}
We would like to thank the Data Agency and Security (DAS) Lab at George Mason University and Coventry University where the study was conducted. The opinions expressed in this work are solely those of the authors.


\bibliographystyle{IEEEtran}
\bibliography{citations}



\section*{Supplementary material}

# Privacy-Preserving Compliance Checks on Ethereum via Selective Disclosure

**Supriya Khadka[1], Dhiman Goswami[2], Sanchari Das[2]**
[1] Coventry University, Coventry, United Kingdom
[2] George Mason University, Fairfax, Virginia, United States

## Abstract

Digital identity verification often forces a privacy trade-off, where users must disclose sensitive personal data to prove simple eligibility criteria. This work proposes a Selective Disclosure Framework on Ethereum designed to decouple attribute verification from identity revelation. By utilizing client-side zk-SNARKs, the framework enables users to prove specific eligibility predicates (e.g., Age > 18) without revealing underlying identity documents. We present a case study, ZK-Compliance, which implements a functional *Grant, Verify, Revoke* lifecycle. Preliminary results indicate that strict compliance requirements can be satisfied with negligible client-side latency (< 200 ms) while preserving the pseudonymous nature of public blockchains.

## Introduction

**Conflict**: Public blockchains require radical transparency, creating hostile environment for (PII) [1,2].

**Trap**: Current KYC paradigms force users to link real-world identities to immutable transaction histories, destroying pseudonymity [3-5].

**Solution**: Selective Disclosure Framework using client-side zk-SNARKs [6] to decouple attribute verification (e.g.,Age >18) from identity revelation.

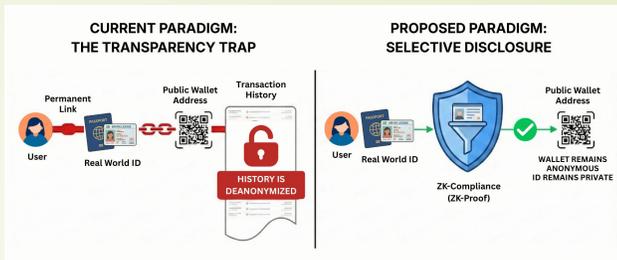

**Figure 1**: The Transparency Trap vs. Privacy-Preserving Selective Disclosure.

## System Architecture

**Identity Vault (Client)**: Encrypted local storage for raw attributes

**ZK Prover (Browser)**: A SnarkJS module that generates proofs locally using private inputs and a cryptographic salt.

**Verifier Contract:** On-chain logic that validates cryptographic proofs without viewing inputs.

**Access Registry:** A state manager that records valid access grants and handles revocation.

## The "Grant, Verify, Revoke" Lifecycle

**Grant (Proof Generation)**: User inputs attribute and salt. Client generates a zk-SNARK proof (< 200 ms) and submits it to the Access Registry.

**Verify (Access Consumption)**: DApp queries Registry. Verification is constant-time lookup of AccessRecord, avoiding heavy on-chain cryptography.

**Revoke (Sovereign Control)**: User can trigger a "Kill Switch" transaction at any time, instantly deleting AccessRecord and restoring privacy boundaries.

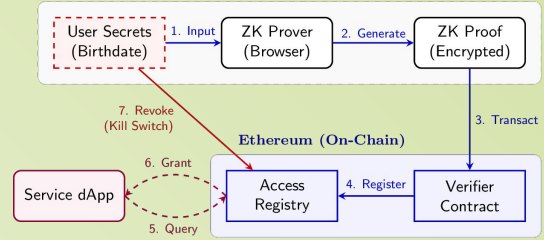

**Figure 2**: The Grant, Verify, Revoke lifecycle. (1-4) The user grants access via a ZK proof. (5-6) The dApp consumes the access. (7) The user can revoke permission at any time.

## Case Study: ZK-Compliance and Results

**Implementation Stack**
- **Circuit**: Circom (Range Check + Salt Hashing)
- **Prover**: SnarkJS (Browser-based)
- **Network**: Ethereum Sepolia Testnet (BN128 Curve)

**Performance Metrics**
- **User Experience:** Client-side proof generation takes < 200 ms on consumer hardware.
- **Gas Costs**: Compared the cost of the on-chain Groth16 pairing check (~240,512 gas) across different environments:

| Deployment | Cost (Est.) | Status |
| --- | --- | --- |
| Ethereum Mainnet | ~$15.00 | Infeasible |
| Layer 2 (e.g.,Arbitrum) | <$0.50 | Viable |

## Conclusion and Future Work

**Conclusion**
Regulatory compliance on public blockchains doesn't require the *transparency trap* of over-disclosure. By using client-side zk-SNARKs, the proposed framework allows users to satisfy strict eligibility predicates while maintaining pseudonymity and sovereign control.

**Future Work**
- Cryptographically binding proofs to the user's msg.sender address to prevent mempool theft.
- Implementing dApp-scoped *nullifiers* to prevent de-anonymization via cross-application clustering.
- Investigating the isolation of proving logic within secure enclaves.



\end{document}